\documentstyle[aps,pre,preprint]{revtex}

\newcommand{\be}{\begin{equation}}
\newcommand{\ee}{\end{equation}}
\newcommand{\ba}{\begin{eqnarray}}
\newcommand{\ea}{\end{eqnarray}}

\begin{document}
\bibliographystyle{unsrt}
\draft

\title{
Moving discrete breathers?
}

\author{S. Flach and K. Kladko }

\address{Max-Planck-Institute for Physics of Complex Systems, Bayreuther
Str. 40 H.16, D-01187 Dresden, Germany}

\date{\today}

\maketitle

\begin{abstract} 
We give definitions for different types of moving
spatially localized objects in discrete nonlinear lattices.
We derive general  analytical relations connecting frequency, velocity and localization length
of moving discrete breathers and kinks in nonlinear one-dimensional lattices. 
Then we propose
numerical algorithms to find these solutions.  
\end{abstract}

\section{Introduction}

The search for moving radiationless spatially localized excitations 
is 
an interesting topic  in the field of nonlinear dynamics of
systems of many interacting degrees of freedom.

Many of the integrable
one-dimensional models, both discrete and continuous in space, 
posess moving objects,
either breathers or kinks \cite{degm82}. 
The integrability of these models provides one with
action-angle variables. Tuning the actions as parameters one can continuously 
go from stationary solutions to moving ones (examples are the nonlinear
Schr\"odinger equation \cite{degm82}, the Toda \cite{mt89} 
and Ablowitz-Ladik lattices \cite{al76}). 
This fact allows us to think about
the whole family of solutions as of a particle-like entity. Still one
has to admit that the precise connection between integrability and possible
existence of moving localized objects is not known. The reason is the hidden
character of the symmetries which provide integrability.

Another  reason for the existence of moving solutions 
can be some continuous symmetry of the Hamiltonian, e.g. the invariance
under Lorentz transformations. 
The Lorentz transformation generates moving objects provided
the corresponding stationary object exists. 
The system does not need to be integrable,
so moving kinks exist for instance in $\Phi^4$ theory in 1+1 dimensions
\cite{degm82}. 
However  stationary breathers
appear to be nongeneric for space-continuous models already in
the case of one spatial dimension \cite{bb94-2},\cite{jd93}
The reason for
that  lies in the fact that the phonon frequency spectrum $\Omega_q$ ($q$
is a wave vector)
of small-amplitude (linearized) vibrations of a continuous system 
is typically unbounded from above. Per definition a stationary breather
is a time-periodic spatially localized solution of the equations of motion.
Expanding the solution in a Fourier series with respect to time
one has to deal with the Fourier components associated to each
multiple $k\Omega_b$ of the breather frequency $\Omega_b$ with integer $k$. 
The unboundeness of the phonon spectrum leads to an 
infinite number of high frequency resonances, which 
generally prevents from the appearance of space-localized
time-periodic solutions \cite{ekns84}. 
An exception is e.g. the sine-Gordon system and
some isolated perturbations of it.

Space-discrete models can generically allow  for the 
existence of stationary discrete breathers \cite{sf94},\cite{ma94} 
(see \cite{fw97} for
an extensive discussion and \cite{lit-list} for a list of 
references). One reason for that is
that the discreteness of space produces a lower cutoff in the wavelength
of small amplitude plane waves, and thus a finite upper bound for the
phonon spectrum. Then one has the possibility to choose frequencies 
$\Omega_b$ such that $k\Omega_b \neq \Omega_q$ for all integer $k$.
However discreteness in space implies the loss of say the continuous
Lorentz symmetry. Thus in general there is no clear way how to 
generate moving breathers out of stationary ones.
            
To look for moving breather solutions we need to have some 
good definition of them.
We define the simplest type of a moving solution as a solution that
repeats 
itself after the time $T_s$ shifted  by one lattice site. Such a solution 
is a fixed point of the map $R G_{T_s}$ 
where $G_t$ is the evolution operator in the phase
space of  the system, $R$ is the translation operator that  
shifts all indices by 1.
A little bit more sophisticated but still simple solutions can be obtained by
considering fixpoints of the map $R^n G$, i. e. solutions that repeat themselves
after the  time $T_s$ shifted by $n$ sites. 
We assume the lattice spacing to be 1 so the
velocity $V$ is then just $n/T_s$. 

Depending on the boundary conditions at infinity 
we speak about moving breathers or kinks.
A trivial example of a radiationless moving discrete kink 
can be obtained by considering 
identical  billiard balls on the line separated by some distance $l$. 
Then kicking one ball we get  
an eternal motion where after the n-th kick the n-th ball 
transfers all its energy
to the n+1-th one. The interaction potential of balls can 
be more or less arbitrary the only restriction
is that it is short-ranged enough, so only  two balls interact 
at each moment of time. 
Below we will in general discuss boundary conditions corresponding
to moving breathers, i.e. the lattice is asymptotically in the
same groundstate no matter what direction from the center is chosen,
given only that the distance from the center becomes infinitely large.

In the next section we will outline necessary conditions o fexistence
of moving breathers. In section III examples are given, and numerical
calculations of solutions are presented in section IV. In section V
we relate our results to the work of others.

\section{Necessary conditions of existence of moving breathers}

Since a stationary breather is characterized by an internal frequency
$\Omega_b$, we have to incorporate this timescale into the definition
of a moving breather.
Consider a one-dimensional lattice, describing the interaction of
degrees of freedom associated to each lattice site. Each degree of freedom
is given by a pair of canonically conjugated variables (e.g. displacement
and momentum) labeled with the site index. Call one of those variables
$u_n(t)$.
We define a one-frequency discrete moving breather solution as 
\begin{equation}
u_n (t) = F ( \Omega_b t , n - V t )\;\;. \label{1}
\end{equation}
Here $F(x,y)$  is a function  period $2 \pi $ periodic 
with respect to $x$ and localized with respect to $y$:
\begin{equation}
F(x+2\pi,y)=F(x,y)\;\;,\;\;F(x,y \rightarrow \pm \infty) \rightarrow 0
\;\;. \label{2}  
\end{equation}
If $T_s$ and $2 \pi /  \Omega_b $ are commensurate so $k T_s = 
l 2 \pi/ \Omega_b$, where $k$ and $l$ are integers,  then  
such a breather repeats itself after
time $k T_s$ shifted by $k$ sites and 
belongs to  the simplest moving breathers defined above. In the general 
noncommensurate case the breather will never 
repeat itself although coming arbitrary close to it.

In the same manner breathers having two or $N$ internal 
frequencies can be defined. This hierarchy
of objects incorporates everything that  we 
intuitively  percept as an object moving through the lattice.

Thinking of moving breathers in terms of fixed points allows to define 
other interesting objects on a discrete lattice.
Consider a fixed point of some general map $G_{T_s} X$ 
where $X$ is an element of the lattice symmetry 
group. If $X$ is the identical transformation we get 
stationary breathers. The translation operator gives us
moving ones. For one-dimensional lattice the only 
symmetry group element left is the reflection, which
gives us reflector-breathers, which mirror themselves 
after time $T_s$. Higher dimensional lattices provide
more choices namely taking a rotation as $X$  we get 
rotation-breathers\footnote{
Not to be confused with rotobreathers, see 
\cite{tp96},\cite{ma96},\cite{tp97},\cite{sa97}. }
, and then taking as $X$ a superposition
of rotation and translation we get "walking" breathers. 

It is clear from  the beginning that looking for 
discrete breathers in terms of exact analytical solutions 
is a hard task. Approximate methods like a 
rotating wave approximation  even if justifiable would turn us away from
the phase space of the system and thus 
do not help too much to
understand what a moving discrete breather is. 
A productive way is to look at moving breathers
from a general point of view and although not 
solving any particular problem  to exactly find
model independent relations that all the moving breathers should fulfil.

We derive these relations below considering the 
tails of a moving breather where the motion
can be considered linear because of the small amplitude of oscillations. 

Let us consider a moving breather with one internal frequency
as  defined above in (\ref{1}).
We write the function   $F(x,y)$ in a Fourier series with respect to $x$ :
\begin{equation}
F(x,y) = \sum_k e^{i k x} f_k (y) \;\;. \label{3} 
\end{equation}
Inserting this ansatz into the Hamiltonian equations of motion 
we obtain
coupled equations for the functions $f_k(y)$.
In the spatial tails of the breather these equations decouple with
respect to the label $k$.  
Let us consider some particular value of $k$ 
with frequency $\Omega_k = k \Omega_b$. The equations
are linear so we seek for a solution in exponential form: 
\begin{equation}
f_k(y) = e^{- i q_k y}\;\;. \label{4}
\end{equation} 
Here $q_k$ is a complex number.
The  $u_n(t)$ term corresponding to a given $k$-th 
harmonic then takes the following form: 
\begin{equation} 
u_{nk}(t) = e^{ i (z t - q_k n)}\;\;. \label{5}  
\end{equation}
Here $z= \Omega + V q_k$.
Using the equations of motion we  finally obtain
\begin{equation}
z  = G ( q_k ) \;\;,  \label{dispersion}
\end{equation}
where $G(q)=\Omega_q$ is the dispersion relation of 
the system analytically continued to the
complex plane.

With the definition of $z$ we have
\be
\Im{ z}  =  V  \Im{ q_k}\;\;, \label{main}
\ee
where $(\Im{ q_k})^{-1} = \lambda $ is the localization length 
of a given harmonic, $\Re {z} = \Omega + V \Re {q_k}$,
$\Omega = k \Omega_b$.

This equation connects the frequency,  
velocity and the localization length of a given harmonic with each
other. 
For the particular case of a breather without internal frequency 
$\Omega_b=0$ (shock wave) we obtain the
following relation between the velocity and the localization length:
\be
V  q_k = G(q_k)\;\;. \label{8} 
\ee
This equation shows that for a shock wave with a given velocity 
only a discrete set of complex
wave vectors $q_k$ is allowed. 
For the linear dispersion relation $G(q)= C q$ only the velocity
$V=C$ is allowed and then $q$ can be arbitrary. 
For the parabolic dispersion relation $G(q) = Aq^2$ $q$  
is  always real namely $q= V/A$ so no localization 
is possible because the localization   
length is the inverse of the imaginary part of $q$. 
For the quadratic dispersion relation with a cutoff at finite
frequency
\begin{equation}
V q = G(q) = A q^2 + D\;\;\label{9}
\end{equation}
we have a complex root when  $(V)^2- 4AD < 0$. 
Therefore  a cutoff in the dispersion relation at finite frequencies
in the dispersion relation is a 
necessary condition to have localization of a moving solution.
Note that such a cutoff can be caused either by a gap in a space-continuous
system or simply by a space-discrete system.

For moving breathers with nonzero internal frequency a necessary 
condition of existence is obviously that
all harmonics $f_k(y)$ are localized namely for any $\Omega = k \Omega_b $ 
there is a solution
of equations (\ref{main}) having a complex $q_k$.

Therefore we obtain that for any nonlinear 
lattice the frequency of the breather, its velocity
and the localization length of its harmonics are not 
independent but connected by the equation (\ref{main}).

\section{Examples}

In this chapter we will consider examples of systems to illustrate
the general considerations from above.

\subsection{Space-continuous models}

\subsubsection{Klein-Gordon models in 1+1 dimensions}

Consider the partial differential equation for the field $U(x,t)$
\be
U_{,tt}=-U+CU_{,xx} - F_{nl}(U)\;\;,\label{2-1}
\ee
where the function $F_{nl}(x)$ if expanded in a Taylor series around
$x=0$ contains only nonlinear terms in $x$. Examples are e.g. the
$\Phi^4$ Klein-Gordon equation with $F_{nl}(x)=x^3$ or the sine-Gordon
equation with $F_{nl}(x)={\rm sin}(x)-x$.

Let us search for a moving solution in the form
\be
U(x,t)=\sum_k u_k(x-Vt) {\rm e}^{ik\Omega_b t}\;\;. \label{2-2}
\ee
Inserting (\ref{2-2}) into (\ref{2-1}), assuming that the amplitude
of the solution is small at large distances from some center, skipping
the nonlinear terms in (\ref{2-1}) we obtain for each integer $k$
\be
(V^2-C)\frac{{\rm d} ^2 u_k}{{\rm d} z^2} 
- 2iVk\Omega_b\frac{{\rm d} u_k}{{\rm d}z } +(1-k^2\Omega_b^2)u=0\;\;.\label{2-3}
\ee
For simplicity we skip the $k$ index and define $u=u_k$, $\Omega=k\Omega_b$,
$z=x-Vt$.
To solve (\ref{2-3}) we make the ansatz $u(z)\sim {\rm e}^{\lambda z}$.
Note that $\lambda=iq_k$ (see (\ref{4})). Decomposing $\lambda$ into
real and imaginary parts $\lambda = R + iI$ ($R,I$ real) we find
\ba
(V^2-C)(R^2-I^2) + 2V\Omega I +1-\Omega^2=0 \;\;\label{2-4} \\
(V^2-C)2RI - 2V\Omega R = 0 \;\;. \label{2-5}
\ea
The physically relevant parameters of our solution are the velocity
$V$ and the exponent of the spatial decay  $R$ characterizing the
localization length of the object. The physical frequency describing
the true oscillations is given by $\omega=\Omega - VI$. 
Solving (\ref{2-4}) we find
\ba
I^2=\frac{V^2}{C} \left[ \frac{1}{C-V^2} - R^2 \right]\;\;,\label{2-6} \\
\omega = -\frac{C}{V}I \;\;. \label{2-7}
\ea
It follows that $V^2 < C$ (Lorentz invariance of (\ref{2-1}) and
$V^2 \geq C - 1/R^2$. These two curves define the allowed region in
the $\{R^2,V^2\}$ plane of possible solutions, which is shown in Fig.1.
The velocity of any moving object has to be below the speed of light.
For large (weakly localized) objects with $R^2 < 1/C$ the solution
will always have some nonzero frequency $\omega$, i.e. there is a
gap in the allowed frequency spectrum containing $\omega=0$.
For small (strongly localized) objects with $R^2 \geq 1/C$ the gap closes,
and one can always design a tail solution with zero frequency $\omega=0$
(this corresponds to the lower bound of the allowed region in Fig.1).
The case of zero frequency is nothing but a tail (or front) of a shock wave.
Note that there are no restrictions with respect to the value of $R^2$,
so the tail solutions can be infintely stronlgy localized in space.

So far discussed the solutions without checking whether the initial
ansatz (\ref{2-2}) can be completely fulfilled in the tails, i.e. for
all integers $k$. Note that in ansatz (\ref{2-2}) we parametrize the
solution using $\Omega_b$ and $V$. To answer that question, we can
argue in the following way. Suppose we choose a point in the $\{R^2,V^2\}$
diagram in the allowed region. This point gives us a set of values for
$\omega$ and $I$. Suppose that $k=1$. Then we obtain some unique value
for $\Omega_b$. Now we can go the inverse way and say, that for that
value of $\Omega_b$ and $V$ we find the corresponding values for $R$ and
$I$. But what now for other values of $k$? Since increasing $k$ we increase
$\omega$, we have to check whether at a fixed value of $V$ we can realize
any value for $\omega$ in Fig.1 by changing $R$ throughout the allowed
region. The answer is no. Indeed fixing $V$ we always obtain a finite
line segment in Fig.1. On the right end of this segment $\omega=0$, 
and on the left $\omega < \infty$, with no singularities in between.
Thus we can never realize ansatz (\ref{2-2}). So we conclude that in
general moving breathers do not exist in Klein Gordon field problems.
The only possiblity is to have a shock wave (or kink), i.e. to
set $\Omega_b=I=\omega=0$, which is possible. 

It seems that we are in conflict with the well known result, that
stationary and moving breathers exist in the sine-Gordon PDE. But keep
in mind, that first the sine Gordon PDE is integrable. Secondly if
one takes the stationary breather solution for that case and expands
it in a Fourier series with respect to time, inserts the sum into
the equations of motion and looks at the behaviour of the Fourier
components in the spatial tails of the solution, it appears that
their decay is not described by the linearized equations of motion.
In this nongeneric (since integrable) case some nonlinear terms in
the equations of motion have to be kept to explain the final exponential
decay of breather harmonics which resonate with the phonon band. 
This corresponds to the following: suppose you solve a differential
equation with inhomogeneous terms. Then the full solution is a
sum of the general homogeneous solution and a particular inhomogeneous
solution. In our context the homogenenous part comes from the
linear terms, and the inhomogeneous from some nonlinear terms.
The high symmetry of the sine Gordon equation miracuously ensures
the vanishing of the coefficients in front of the (nondecaying)
homogeneous part. However as already discussed most perturbations
of the sine Gordon equation, by destroying those hidden symmetries,
destroy the breather. So although this is a subtle point, we don't
see any contradiction to our generic statement.

\subsubsection{(Non)linear Schr\"odinger equation}

Consider the partial differential equation for the complex field
$\Psi(x,t)$
\be
\dot{\Psi} = i(C\Psi_{,xx} + F(\Psi))\;\;. \label{2-8}
\ee
Again we search for a solution in the form $\Psi(x,t)=\phi(x-Vt){\rm e}
^{i\Omega t}$. Repeating the same procedures as in the previous case
we arrive at the equations
\ba
I = \frac{V}{2C} \;\;, \label{2-9} \\
\omega = C\left( R^2 - I^2\right) \;\;. \label{2-10}
\ea
Since (\ref{2-8}) is not Lorentz-invariant, we do not find restrictions
on the choice of the velocity $V$. In fact we find no restrictions at all,
the whole parameter space $\{R,V\}$ is allowed. 

\subsection{Space-discrete models}

\subsubsection{Discrete (non)linear Schr\"odinger equation}

The equations of motion are given by
\be
\dot{\Psi}_n = i\left[ |\Psi_n|^{\mu}\Psi_n 
+ C(\Psi_{n-1} + \Psi_{n+1}) \right]\;\;. \label{2-11}
\ee
Here the nonlinear term is characterized by some $\mu > 0$.
Again we search for a moving solution in the form 
$\Psi_n = \phi(n-Vt){\rm e}^{i\Omega t}$. Due to the nonlocality of
the difference operator (as compared to the differential one) the
differential equation for $\phi(z)$ contains now retarded and advanced terms
(the tail is again only considered, nonlinear terms are neglected):
\be
-V \phi'(z) = i\left[ -\Omega \phi(z) + C(\phi(z+1) + \phi(z-1))\right]
\;\;. \label{2-12-1}
\ee
Skipping the intermediate calculations we arrive at the result
\ba
I = {\rm arcsin}\left[ \frac{VR}{2\;{\rm sinh}R} \right]\;\;,
\label{2-12} \\
\omega = 2\;{\rm cosh}R \;{\rm cos}I\;\;. \label{2-13}
\ea
A necessary condition is thus $V \leq 2{\rm sinh}(R)/R$ and is shown in
Fig.2. 
In the region of possible solutions for each pair of $\{R,V\}$ we
now have two solutions due to the periodicity of (\ref{2-12}),(\ref{2-13})
in $I$ - to each solution with a given value of $I_1=I$ we can construct
a second solution with $I_2=\pi-I$.
For any given value of $R$ there is a finite upper bound on
the value of $V$. However with increasing $R$ the threshold value of
the upper bound of $V$ also increases. Again we can find infinitely strongly
localized
and infinitely fast moving tails.

\subsubsection{Klein-Gordon chains}

These models describe the dynamics of atoms on a substrate
and interacting with each other:
\be
\ddot{u}_n = -\alpha u_n - C(2u_n - u_{n-1} - u_{n+1}) + F_{nl}(u_n)
\;\;. \label{2-14}
\ee
As in the field case we search for a moving solution in the form
\be
u_n(t) = \sum_k A_k(n-Vt){\rm e}^{ik\Omega_b t}\;\;.\label{2-15}
\ee
In the tails of the assumed existing solution we obtain
\be
V^2\frac{{\rm d^2} A_k(z)}{{\rm d}z^2} -2{\rm i} k\Omega_b V 
\frac{{\rm d}A_k(z)}{{\rm d}z} = (\Omega^2 -\alpha -2C)A_k(z)
+ C(A_k(z+1) + A_k(z-1)) \;\;. \label{2-16-2}
\ee
Repeating the intermediate calculations as above we arrive at the following
equations (note that we skip the index $k$, so below $\omega=k\Omega_b -
VI$):
\ba
\omega=-\frac{C}{VR}\;{\rm sinh}R \;{\rm sin}I \;\;, \label{2-16} \\
\left[\frac{C}{VR}\right]^2 {\rm sinh^2}R \;{\rm sin^2}I =
V^2R^2 + \alpha +2C(1-{\rm cosh}R\;{\rm cos}I)\;\;. \label{2-17}
\ea
The allowed region in the parameter space $\{R,V^2\}$ is similar to
the one of the discrete nonlinear Schr\"odinger case (Fig.2) but also
more complicated. Solutions exist below a certain line $V^2(R)$. This
line consists of two parts. For ${\rm cosh}R > 2C/(\alpha+2C-
\sqrt{\alpha^2 +4\alpha C})$ the line is given by
\be
(a)\;\;:\;\; V^2=\frac{1}{R^2}(2C{\rm cosh}R - \alpha -2C)\;\;.
\label{2-18}
\ee
For smaller values of $R$ we have line (b):
\be
(b) \;\;: \;\; V^2=\frac{1}{2} \frac{{\rm sinh^2}R}{R^2}
(\alpha + 2C - \sqrt{\alpha^2 + 4\alpha C})\;\;. \label{2-19}
\ee
In this second case the continuation of line (a) (\ref{2-18})
separates solutions with multiplicity 4 (to the left) from solutions
with multiplicity 2 (to the right) as shown in Fig.3.

Let us fix any value of $V$. This value defines us some half-infinite
line segment of allowed solutions in $\{R,V^2\}$. For $\omega$ we
find $0 \leq \omega < \infty$. Thus for any $V\neq 0$ we can generate
a breather solution in the tails, for any value of $\Omega_b$! Surprisingly
the problem of resonances, as in the case of a stationary breather,
does not appear on that stage. 

We can also find a solution to $\Omega_b=\omega=I=0$, i.e. we can
again generate shock waves (or kinks) with any velocity (in the tails).
They correspond to solutions on line (a).

\subsubsection{Acoustic chains}

This case is obtained by performing the limit $\alpha \rightarrow 0$.
The phonon spectrum is acoustic. In that case solutions exist below
line (a) (\ref{2-18}), which now extends down to $R=0$ (Fig.4). All solutions
are of multiplicity 2. In contrast to the previous case for velocities
$V^2 < C$ we can generate moving breather tails only for frequencies
above some threshold value. This gap value shrinks to zero and remains
zero as the velocity is increased above $V^2 > C$. Also shock waves
(kinks) which correspond to all points on line (a) can be again generated.
Again in contrast to the previous case, these shock waves or kinks
must have velocities $V^2 > C$ to be generated.

\subsection{Some discussions}

Let us summarize the results obtained so far. We assumed the existence
of some moving localized object (breather) which is parametrized in a proper
way. Then we considered the equations of motion in the tails of the object,
and checked under what conditions the linearized equations in the tails
can be satisfied. We found, that Klein Gordon PDEs do not allow in general
for moving breathers (as they do not for stationary ones), only shock
waves or kinks are allowed. Discretizing these equations, we found
that all restrictions are gone, and moving breather tails can be
generated for any parameters. It is surprising that even the nonresonance
conditions known to exist for stationary breathers do not appear here.
If we consider a chain with an acoustic spectrum, then the nonresonance
condition reappears in some sense, but the forbidden frequency gap shrinks
to zero as the velocity is increased above the speed of sound.

Another result is, that we can generate shock wave tails in all cases,
with restrictions in some cases (acoustic chains) on the velocity.

\section{Numerical methods}

\subsection{The general idea}

Let us consider a chain with
the equations of motion given by $\ddot{u}_n =- \partial H /
\partial u_n$. Then the ansatz (\ref{2-15}) yields equations
of the type (cf. (\ref{2-16-2}))
\be
V^2\frac{{\rm d^2} A_k(z)}{{\rm d}z^2} -2{\rm i} k\Omega_b V
\frac{{\rm d}A_k(z)}{{\rm d}z} =
\Omega^2 A_k(z) + \sum_n f_n(\{A_{k'}(z)\},\{A_{k'}(z+n)\},\{A_{k'}(z-n)\})
\;\;. \label{3-1}
\ee
The essential feature is that these coupled differential equations
contain advanced and retarded terms. 
These terms arise due to the interaction on the lattice.
Instead of directly trying to
solve these equations, we consider a lattice governed by the equations
\be
V^2 \ddot{A}_{kn}(t) -2{\rm i} k\Omega_b V \dot{A}_{kn}(t) =
\Omega^2 A_{kn}(t) + \sum_{n'}f_{n'}(\{A_{k'n}(t)\},
\{A_{k',n+n'}(t)\},\{A_{k',n-n'}
(t)\})
\;\;. \label{3-2}
\ee
Here $n$ is again the lattice site label, and with each lattice site $n$
we have an associated infinite set of variables $\{A_{kn}\}$, $
k=0,\pm 1, \pm 2, ...$. Equations (\ref{3-2}) define a phase space flow
in the phase space of all variables $A_{kn},\dot{A}_{kn}$. In general
trajectories generated by that dynamics are not related to solutions
of (\ref{3-1}). However all fixed points of the map $RG_{t=1}$ ($G_t$ is
the evolution operator defined by (\ref{3-2}) and $R$ the translation
operator that shifts all lattice indices by 1) are solutions of (\ref{3-1}).
The main reason for that is that all delay and advance intervals are 
integers.

Once a fixed point (solution) is found, it can be continued using
generalized Newton methods or steepest descent methods.

\subsection{Example: DNLS}

Let us investigate numerically moving breathers in
the discrete nonlinear Schroedinger model (DNLS) with $\mu=2$ (\ref{2-11})  
which is a nonintegrable
model.

The DNLS system has an integrable counterpart which is the Ablowitz-Ladik
model (ALM) having the following Hamiltonian

\begin{equation}
\dot{\Psi}_n = 
{\rm i} (\Psi_{n-1} + \Psi_{n+1})(1 +|\Psi_n|^2) \;\;.
\label{4-1}
\end{equation}

The  ALM has moving breathers of the form:

\begin{equation}
\Psi_n = \frac{\sinh (\mu)}{\cosh (\mu (n-Vt))}{\rm e}^{-ikn}
{\rm e}^{i\Omega_b t}  \;\;.  
\label{4-2}
\end{equation}

Here $k$ and $\mu$ are free parameters.

All other parameters are expressed through $k$ and $\mu$

\begin{equation}
\label{4-3}
\Omega_b= 2 \cosh (\mu)\cos (k)\;\;,
\end{equation}

\begin{equation}
\label{4-4}
V= \frac{2}{\mu}\sinh (\mu)\sin (k) \;\;.
\end{equation}
Here $V$ is the velocity of a breather. 

Now we can test the relations obtained in the previous section. 
Indeed, using
the solution (\ref{4-2}) in the tails it can be  checked that the 
relation between the frequency, velocity and localization length 
(\ref{2-12}),(\ref{2-13}) is fulfilled.
Note that in this case we have only one Fourier component which is
a specific property of models of the  nonlinear Schroedinger type which 
do not generate higher Fourier harmonics with respect to time
if a single Fourier component is excited.

Now let us consider the following model, which allows for a continuous tuning
between ALM and DNLS:
\begin{equation}
\label{4-5}
\dot{\Psi}_n ={\rm i}\left[ \Psi_{n-1} + \Psi_{n+1} 
+|\Psi_n|^2 [(1-\alpha)(\Psi_{n-1}+\Psi_{n+1}
)+\alpha \Psi_n]
\right]
\;\;.
\end{equation}

Here $0 \le \alpha \le 1 $.

Let us look for a solution in the form
\begin{equation}
\label{4-6}
\Psi_n (t) = {\rm e}^{i\Omega_b t} g_n (t)\;\;.
\end{equation}

Then we have
\begin{equation}
\label{4-7}
\dot{g}_n = 
{\rm i} \left[g_{n-1} + g_{n+1} - 2 \Omega_b g_{n} 
+|g_n|^2 [(1-\alpha)(g_{n-1}+g_{n+1}
)+\alpha g_n]
\right]
\;\;. 
\end{equation}

For a moving breather we have 
\begin{equation}
g_n(t) = g(n-Vt)\;\;.
\end{equation}

The moving breather is a fixed point of the map $  R G_T$, $T= 1/V$. We look for a zero minimum
of the functional $F= |RG_T X -X|$ 
where $X$ is a point in the phase space of (\ref{4-7}).

We proceed in the following way. First
the parameter $\alpha$ is put to zero. 
The initial point in the phase space is chosen to be 
the ALM moving breather (\ref{4-2}).
Then $\alpha$ is incremented by a small value $\Delta \alpha$.
A minimization of the functional $F$ is performed. Again 
$\alpha$ is incremented by a small value $\Delta \alpha$, etc.
The algorithm stops when $\alpha$ reaches $1$.

We were able to generate 
moving DNLS breathers with the value of the 
minimized functional  less then $10^{-6}$. 
An example is shown in Fig.5 ($\mu=0.5$, $k=1$, $V=0.0364$).

Let us mention an important  point found during the numerical calculations
which is the existence of a large (infinite?) 
number of very (infinitesimally?) close local nonzero minima 
of the functional $F$ near the true fixed point.
When a step $\Delta \alpha$ is made, 
the algorithm minimizes the functional $F$, and we hope that
the value of the functional in the minimum will be zero.
Surprisingly (and in contrast to the numerical calculations
on stationary breathers) we found 
that the value of the functional at the minimumi for a given
$\alpha$ is not
zero {\it no matter how small $\Delta \alpha$ is}. 

On the other hand when  $\Delta \alpha$ is decreased 
(thereby increasing the computation time) 
the minimum value of the functional at a given
$\alpha$ tends to zero, although not being zero 
for any finite $\Delta \alpha$. From this we conclude 
that the true fixed point trajectory is surrounded
by a dense set of other nonzero minima of the $F$. 
The structure of the phase space near
the moving breather is therefore highly nontrivial 
in contrast to stationary breathers where none 
of above effects were found. 

These findings are maybe connected to the fact, that
the spectrum of Floquet multipliers of a moving breather
has unusual properties as compared to the Floquet spectrum
of stationary breathers. 
Namely, the Floquet multipliers of the linearized map around  
a moving breather 
fill the unit circle densily. Especially there exist
Floquet multipliers with value +1, which would in general
make continuation impossible for stationary breathers. 
The existence of these multipliers can be simply explained.
Linearizing the map around a moving breather fixed point,
we obtain an infinite set of eigenvalues with spatially
extended eigenvectors. At large distances from the breather
center these eigenvectors will correspond to normal phonons.
A phonon is given by 
\be
{\rm e}^{{\rm i}\Omega_qt -qn}\;\;.
\ee
It is always possible to cast it into the form 
\be
{\rm e}^{{\rm i}\Omega_b t} {\rm e}^{-{\rm i}(n-Vt)}
\ee
with arbitrary numbers $\Omega_b,V$ by solving 
\be
\Omega_q = \Omega_b + Vq\;\;.
\ee
Indeed we can always find $q$-values which will do the job.
For the case of a stationary breather $V=0$, and we essentially
recover the nonresonance condition, which can be fulfilled
by choosing $\Omega_b$ to be outside the phonon band.

\section{Conclusion}

In this section we will briefly discuss related work. 

\subsection{Moving kinks}

Moving kinks can be considered to some extend as moving breathers
with zero frequency. What matters here is that these objects can
be represented by one (zero frequency) Fourier component in
(\ref{3}). Proofs of existence of moving kinks in FPU chains
have been obtained in \cite{gfjadw94} by finding them as
minimisers of a variational problem and in \cite{jcjbml96}
by analytical continuation from the integrable Toda lattice.

Numerical solutions for moving kinks have been obtained e.g. in 
\cite{dhfgmhb89} and \cite{jcerf90}. Fourier
transformations in space are used in the first one, while the second
paper treats space-periodic solutions, but does not uses Fourier
transformations.

\subsection{Moving breathers}

There is a large amount of work reporting on moving breathers
in FPU chains (e.g. \cite{ht92-jpsj-1}). A study of
their connection with stationary breathers was started in
\cite{fw94}. In \cite{cat96} this connection was used to numerically obtain
moving breathers by exciting pinning modes of stationary breathers.

Finally moving breathers have been obtained numerically for
DNLS chains in \cite{defw93}.

So far we are not aware of existence proofs for moving breathers.
\\
\\
\\
\\
\\
Acknowledgements
\\
\\
We thank S. Aubry, T. Cretegny, J. C. Eilbeck, 
R. S. MacKay and V. Pokrovsky for useful discussions. 

\newpage

\newpage
\noindent
FIGURE CAPTIONS
\\
\\
\\
Fig.1. Phase diagram for necessary conditions on the existence
of moving localized objects in Klein-Gordon models in 1+1 dimensions
(see text).
\\
\\
Fig.2 Same as Fig.1, but for the DNLS chain.
\\
\\
Fig.3 Same as Fig.1 but for a Klein-Gordon chain. The labels M2,M4
indicate the multiplicity of solutions in the given part of the
parameter space.
\\
\\
Fig.4 Same as Fig.1 but for FPU chains.
\\
\\
Fig.5 A moving breather solution for DNLS (see text).

\end{document}